\documentclass[aps,prl,twocolumn,showpacs,superscriptaddress,groupedaddress]{revtex4}  
\usepackage{graphicx}  
\usepackage{dcolumn}   
\usepackage{bm}        
\usepackage{amssymb}   
 \usepackage{slashed}
\usepackage{epsf}
\usepackage{hyperref}
\usepackage{lipsum}
\usepackage{enumerate}

\newcommand{\bea}{\begin{eqnarray}}
\newcommand{\eea}{\end{eqnarray}}
\newcommand{\la}{\label}
\newcommand{\be}{\begin{equation}}
\newcommand{\ee}{\end{equation}}
\newcommand{\tr}{\,\mbox{tr}\,}

\begin{document}

\title{$\mathcal{P}$-Odd Pion Azimuthal Charge Correlations in Heavy Ion Collisions}

\author{Yachao Qian}
\address{Department of Physics and Astronomy,
Stony Brook University,  Stony Brook, NY 11794-3800.}

\author{Ismail Zahed}
\address{Department of Physics and Astronomy,
Stony Brook University,  Stony Brook, NY 11794-3800.}

\date{\today}

\begin{abstract}
We argue that the large instanton induced Pauli form factor in polarized proton-proton scattering  may 
cause, through topological fluctuations, substantial charge-dependent azimuthal correlations 
for $\pi^{\pm}$ production in peripheral heavy ion collisions both at RHIC and LHC, thanks to the
large induced magnetic field. Our results compare favorably to the measured pion azimuthal correlations
by the STAR and ALICE collaborations.
\end{abstract}

\maketitle

\section{\label{sec:introduction}introduction}
Large single spin asymmetries  in dedicated semi-inclusive deep inelastic scattering
were reported by both the CLAS and the HERMES collaborations~\cite{hermes2000,hermes2005,hermes2009,clas2010}.
Similarly large spin asymmetries were reported by the STAR and PHENIX collaborations 
~\cite{star2008,Eyser:2006bc, fnal1991} in pion production using a polarized
proton beam at collider energies. These large spin asymmetries  are due to 
chirality flip contributions in the scattering amplitude that are not supported by
QCD perturbation theory and factorization.

The QCD vacuum supports large instanton-antinstanton fluctuations that are non-perturbative
in nature and a natural source for chirality flip  effects. QCD instantons are hedgehog in color-spin,
that makes them ideal for triggering large spin asymmetries~\cite{Forte:1989ia,Forte:1990xb, Ioffe:1990xm,   Kochelev:1999nd,
dorokhov2009, ostrovsky2005, Qian:2011ya, Qian:2014}.  These large chirality flip contributions are beyond the realm of
 factorization and provides a QCD based quantitative mechanism for large spin effects in the initial state 
 (Sivers)~\cite{sivers1990, sivers1991}.

In this note we would like to argue that the chirality flips from instanton and anti-instanton 
in polarized proton-proton collisions may cause, through vacuum topological fluctuations, large pion azimuthal correlations 
in peripheral heavy ion collisions, thanks to the
large induced magnetic field in the prompt phase of the collision.
The organization of this note is as follows: we first briefly review the origin of some of the chirality flip effects
in the QCD vacuum. We then argue that in peripheral heavy ion collisions, a substantial magnetic field
could trigger large polarizations in the protons participating in the collisions. These effects lead to large
pion azimuthal correlations that are comparable to those recently reported by the STAR and ALICE
collaborations. Our conclusions follow.

\begin{figure}
\includegraphics[height=45mm]{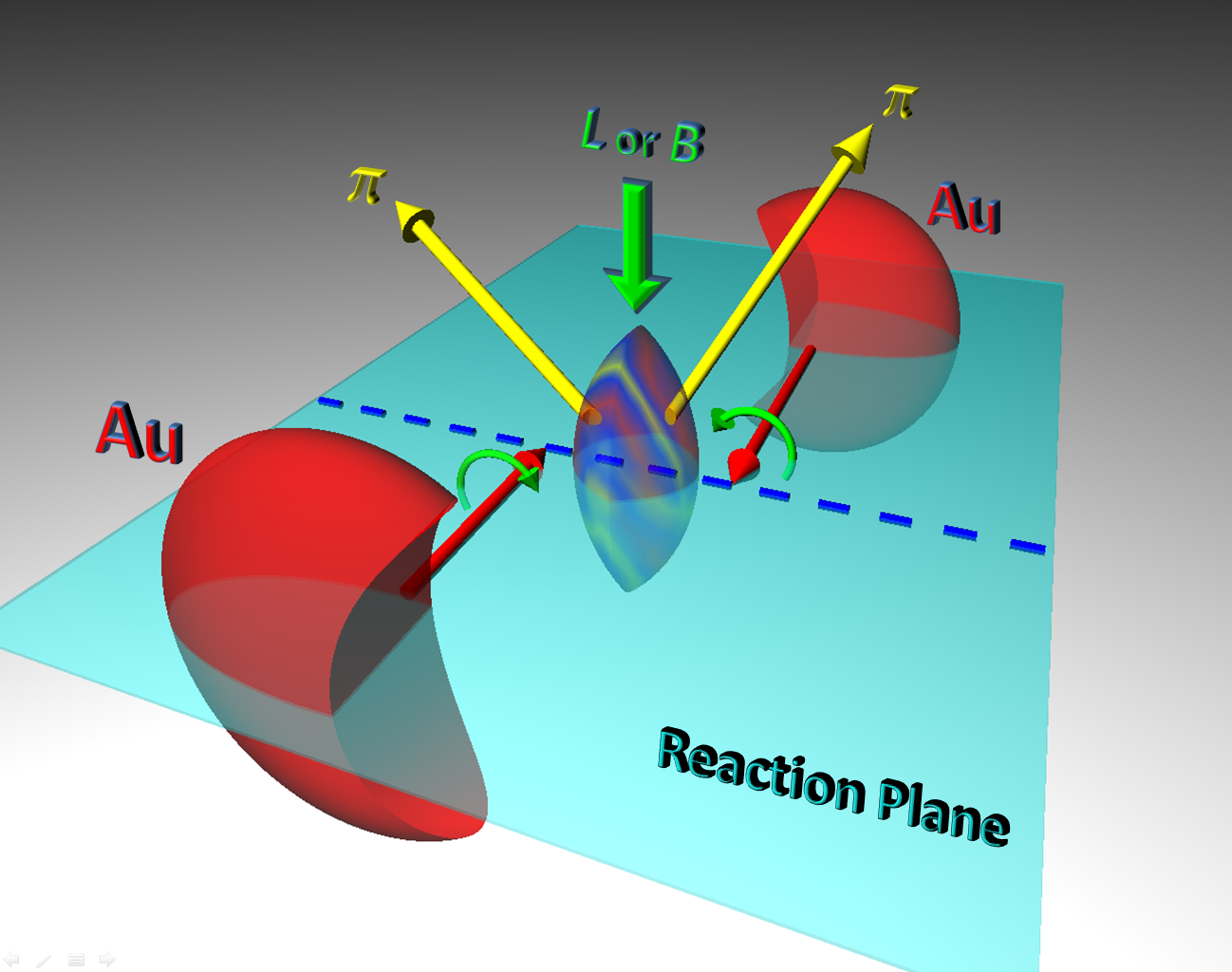}
\caption{2-pion correlations in peripheral  $AuAu$ collisions.}
\label{aacollision}
\end{figure}

\section{P-Odd Effects in the Instanton Vacuum}
\la{sec:podd}

In a typical non-central $AuAu$ collision at RHIC as illustrated in Fig.~\ref{aacollision}, the
flying fragments create a large magnetic field that strongly polarizes the wounded or participant
nucleons. The magnetic field is typically $eB/m_\pi^2\approx 1$ at RHIC and $eB/m_\pi^2\approx 15$
at the LHC lasting for about 1-3 ${\rm fm}/c$~\cite{Skokov:2009qp}. We recall that in these units 
$m_\pi^2\approx 10^{18}\,{\rm Gauss}$ which is substantial and therefore a major source of
prompt proton polarization. Polarized proton on proton scattering can exhibit large chirality flip
effects through instanton and anti-instanton fluctuations as we now show.

\begin{figure}
\includegraphics[height=32mm]{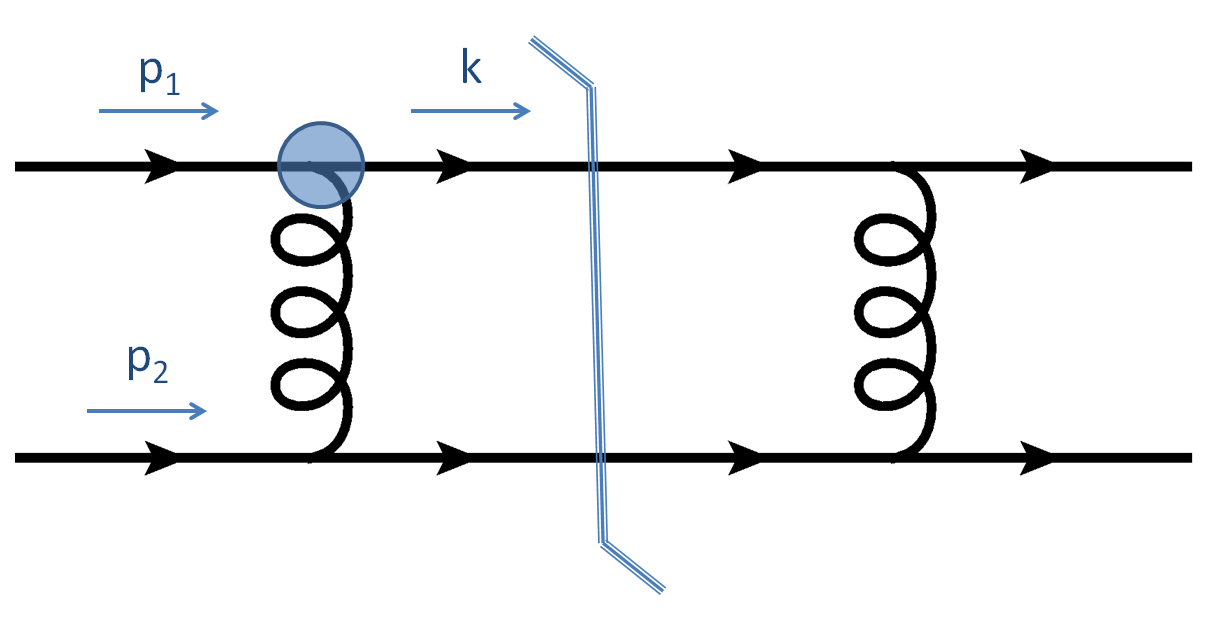}
\caption{ Gluon Exchange. The blob is an instanton or anti-instanton insertion. See text.}
\label{gluonexchange}
\end{figure}

Consider the typical parton-parton scattering amplitude of Fig.\ref{gluonexchange} with 2-gluon exchanges.
In each collision, the colliding "parton" $p_i$ has spin $s_i$, and thus $u(p_i) \bar{u}(p_i) = \frac{1}{2} \slashed{p_i}(1 + \gamma_5 \slashed{s}_i)$. The parton $p_1$ from the A-nucleus encounters an instanton or anti-instanton 
as depicted by the gluonic form-factor. The latter follows from standard instanton calculus~\cite{Kochelev:1996pv}

\be  
\label{VERTEX}
M_\mu^a = t^a  \left[  \gamma_\mu-   \bold{P}_+ \gamma_+    \sigma_{\mu\nu} q^\nu    \Psi     -   { \bold{P}}_- \gamma_-     \sigma_{\mu\nu} q^\nu  \Psi   \right]
\ee
with $\gamma_\pm =(1\pm \gamma_5)/2$ and

\be
\Psi =   \frac{   F_g(\rho_c\, Q)   \pi^4   (n_I \rho^4_c)   }{m_q^* g_s^2}
\ee
and $F_g (x) \equiv {4}/{x^2} - 2 K_2 (x)$ with $F_g(0)=1$. Here $n_I\approx 1/{\rm fm}^4$ is the effective 
instanton density, $\rho_c\approx 1/3\,{\rm fm}$ the typical instanton size and $m_q^*\approx 300\,{\rm Mev}$ the constitutive
quark mass in the instanton vacuum. The momenta of the incoming partons as well
as the momentum $Q$ of the transferred gluon are assumed small or $p\rho_c, Q\rho_c\leq 1$.
$\bold{P}_+ = 1$ stands for an instanton insertion and ${\bold{P}}_-=1$  for an  anti-instanton insertion. 
 In establishing (\ref{VERTEX}),  the instanton and anti-instanton zero modes are assumed to be 
undistorted by the prompt external magnetic field. Specifically, the chromo-magnetic field $B_G$
is much stronger than the electro-magnetic field $B$, i.e. $|g_sB_G|\gg |eB|\approx $ or
$m_\pi^2\rho_c^2\approx 0.004\ll 1$. The deformation of the instanton zero-modes by a strong magnetic field
have been discussed in~\cite{Basar:2011by}. They will not be considered here.

In terms of (\ref{VERTEX}), the contribution of Fig.\ref{gluonexchange} to the differential cross section is

\bea
d \sigma &\sim&  \frac{g_s^4}{ |p_1 - k|^4}    \tr[ M_\mu^a  \slashed{p}_1 (1 + \gamma_5 \slashed{s}_1 )   \gamma_0 (M_\nu^b)^\dagger   \gamma_0 \slashed{k}  ] \nonumber\\
&&\times \tr[\gamma^\mu t_a \slashed{p}_2 (1+\gamma_5 \slashed{s}_2)\gamma^\nu t_b \slashed{k}^\prime]
\eea 
which can be decomposed into $d\sigma\approx d\sigma^{(0)}+d\sigma^{(1)}$  in the dilute instanton liquid.
The zeroth order contribution is

\be \la{zeroorderdiff}
d^{(0)} \sigma  \sim    64 g_s^4   \frac{ 2 (k  \cdot p_2) (p_1 \cdot p_2) +  (k  \cdot p_1) (p_1 \cdot p_2 - k\cdot p_2)  }{ |p_1 - k|^4} 
\ee 
where we used $k^\prime = p_1 + p_2 - k$. The first order contribution is 

\bea
d^{(1)} \sigma &\sim&   \frac{64  g_s^4  }{ |p_1 - k|^4}     \left[  (p_1 \cdot p_2)^2 + (k \cdot p_2) (p_1 \cdot p_2)    \right] \nonumber\\
&&\times(k \cdot s_1) \left(   \bold{P}_+ -   { \bold{P} }_-    \right) \Psi 
\eea
after using $p_1 \cdot s_1 =  0$ and $p_1^2 = k^2 = 0$. Converting to standard parton kinematics with
$p_1 \rightarrow x_1 P_1$, $p_2 \rightarrow x_2 P_2$ and $k \rightarrow {K}/{z}$, we
obtain for the ratio of the $\mathcal{P}$-odd to $\mathcal{P}$-even contributions in the differential cross section

\bea
\frac{d^{(1)} \sigma }{d^{(0)} \sigma } &=&  \frac{ x_1   (P_1 \cdot P_2)^2 + \frac{1}{z} (K \cdot P_2) (P_1 \cdot P_2)   }{ 2    (K  \cdot P_2) (P_1 \cdot P_2) +   (K \cdot P_1) ( \frac{x_1}{x2} P_1 \cdot P_2 -\frac{ K \cdot P_2}{z x2 } )}  \nonumber\\
&&\times   (K \cdot s_1) \left(   \bold{P}_+ -   { \bold{P} }_-    \right) \Psi 
\eea

Now consider the kinematics appropriate for the collision set up in Fig.~\ref{aacollision},
\bea\la{kinematicssimple}
P_{1/2} &=&  \frac{\sqrt{s}}{2} (1, 0, 0, \pm 1) \nonumber\\
K &=& (E ,  K_\perp \cos \Delta\phi, K_\perp \sin \Delta\phi ,  \frac{\sqrt{s}}{2} x_F ) \nonumber\\
s_1 &=&  (0,0, s^\perp_1 , 0)
\eea
where $K_\perp$  and $E ^2= {K_\perp^2 + s x_F^2/4 + m_\pi^2 }$ are  the transverse momentum and total squared energy of the outgoing pion respectively. $x_F$ is the pion longitudinal momentum fraction. Thus

\bea\la{crosssection}
\lim_{s \rightarrow \infty}  \frac{ d^{(1)} \sigma }{d^{(0)} \sigma}  &=&     (\sin \Delta\phi)  s^\perp_1  \frac{x_F + x_1 z}{  x_F z}  \frac{K_\perp}{m_q^*}  \frac{    \pi^3   (n_I \rho^4_c)   }{ 8 \alpha_s} \nonumber\\
&& \times  F_g \left(\rho \sqrt{ \frac{x_1 }{x_F z} (K_\perp^2 + m_\pi^2)}\right) \left(   { \bold{P} }_-  - \bold{P}_+   \right) \nonumber\\
\eea
We note that Eq.~\ref{crosssection} vanishes after averaging over the instanton liquid background which is $\mathcal{P}$-even

\be
\left<  \frac{d^{(1)} \sigma }{d^{(0)} \sigma }  \right> = 0
\ee
since on average $\left< \bold{ Q } \right> = \left<  \bold{P}_+  -   { \bold{P} }_-   \right> = 0$.

\section{P-odd correlations  in AA Collisions through Instantons}
\la{sec:aacollision}

Now consider hard $pp$ collisions in peripheral $AA$ collisions as illustrated in Fig.~\ref{aacollision}. The Magnetic field is strong enough to polarize the colliding protons. For simplicity,  we set $s_\perp u(x,Q^2) = \Delta_s u (x,Q^2)$ and $s_\perp d(x,Q^2) = \Delta_s d (x,Q^2)$, with $\Delta_s u (x,Q^2)$ and $\Delta_s d (x,Q^2)$ as the spin polarized distribution functions of the valence up-quarks and valence down-quarks in the proton respectively. We also assume that the outgoing $u$ quark turns to $\pi^+$ and that the outgoing $d$ quark  turns to $\pi^-$. With this in mind, we may rewrite the ratio of differential contributions
in (\ref{crosssection}) following~\cite{Abelev:2009ac,Abelev:2009ad,collaboration:2011sma,Selyuzhenkov:2012py} as

\be
\frac{d {\bf N}}{ d \phi_{\alpha}} \sim 1 -  2 a_{\alpha} \sin (\phi - \Psi_{RP})
\ee
with $\alpha=\pm $ or 
\be 
a_+  =    \frac{\Delta_s u(x  ,Q^2)}{u(x  , Q^2)}    \Upsilon  \bold{ Q }  \ \ \ \ \ \ \ \ 
a_-  =     \frac{\Delta_s d(x  ,Q^2)}{d(x  , Q^2)}  \Upsilon  \bold{ Q } 
\ee 
and
\be
\Upsilon \equiv \frac{ x_F   +   x     z   }{    x_F    z   }  \frac{K_\perp}{m_q^*}  \frac{    \pi^3   (n_I \rho^4_c)   }{ 16 \alpha_s}  F_g \left(\rho \sqrt{ \frac{  x    }{  x_F  z } (K_\perp^2 + m_\pi^2)}\right)
\ee
While on average $\left<a_\alpha \right>=0$ since $\left<{\bf Q}\right>_V=0$, in general $\left< a_{\alpha} a_{\beta} \right> \neq 0$
for the 2-particle correlations.
Explicitly

\bea
- \left< a_{\pi^+}  a_{\pi^-}  \right> &=& - \left( \frac{\Delta_s u(x  ,Q^2)}{u(x  , Q^2)}  \frac{\Delta_s d(x  ,Q^2)}{d(x  , Q^2)} \right)   \Upsilon^2   \left<  \bold{Q}^2 \right>_V   \nonumber\\ 
-  \left< a_{\pi^+}  a_{\pi^+}  \right>&=& - \left( \frac{\Delta_s u(x  ,Q^2)}{u(x  , Q^2)} \right)^2   \Upsilon^2   \left<  \bold{Q}^2 \right>_V    \nonumber\\
- \left< a_{\pi^-}  a_{\pi^-}  \right> &=& - \left( \frac{\Delta_s d(x  ,Q^2)}{d(x  , Q^2)} \right)^2   \Upsilon^2   \left<  \bold{Q}^2 \right>_V  
\eea
According to \cite{Hirai:2006sr, Adams:1997dp}, $\Delta_s u(x  ,Q^2)/u(x  , Q^2) =  0.959 - 0.588 (1 - x^{1.048})$ and $\Delta_s d(x  ,Q^2)/d(x  , Q^2) = - 0.773 + 0.478 (1- x^{1.243})$. For reasonable values of $\left< x \right>$, $\left< a_{\pi^+}  a_{\pi^+}  \right> \sim \left< a_{\pi^-}  a_{\pi^-}  \right> \sim -\left< a_{\pi^+}  a_{\pi^-}  \right>$ as expected~\cite{Abelev:2009ac,Abelev:2009ad,collaboration:2011sma,Selyuzhenkov:2012py}.

A more quantitative comparison to the reported data in ~\cite{Abelev:2009ac, Selyuzhenkov:2012py} can be carried out by estimating the fluctuations
of the topological charge ${\bf Q}$ in the prompt collision 4-volume $V\approx (\tau^2/2) \Delta\eta V_\perp(b)$. 
In the latter, $\tau\approx$ 1-3 
{\rm fm} is the prompt proper time over which the induced magnetic field is active, $\Delta\eta$ is the interval in
pseudo-rapidity and $V_\perp(b)$ the transverse collision area for fixed impact parameter $b$. Through simple
geometry

\be
V_\perp(b)=2R^2\left({\rm arccos}\left(\frac b{2R}\right)-\frac b{2R}\sqrt{1-\left(\frac b{2R}\right)^2}\right)
\ee
where $R$ is the radius of two identically colliding nuclei. ${\bf Q}^2$ involves a pair ${\bf P}, {{\bf P}^\prime}$ of instanton-antiinstanton.
Specifically,

\be
\label{PAIR}
\left<{\bf Q}^2\right>_V=\left<({\bf P}_+-{\bf P}_-)({\bf P}_+^\prime-{{\bf P}_-^\prime})\right>_V
\ee
If we denote by $N_\pm$ the number of instantons and antinstantons in $V$, 
with $N=N_++N_-$ their total number, then in the instanton vacuum the pair correlation follows from

 \bea
\label{PAIR1}
\left<{\bf Q}^2\right>_V &\equiv& \left<\left(\frac{N_+-N_-}{N_++N_-}\right)^2\right>_V  \approx \frac{\left<(N_+-N_-)^2\right>_V}{\left<(N_++N_-)^2\right>_V} \nonumber\\\
&\approx& \frac{\left<N\right>_V}{\left<N\right>_V
(\left<N\right>_V+4/{\bf b})}
\eea
The deviation from the Poissonian distribution in the variance of the number average reflects on the
QCD trace anomaly in the instanton vacuum  or $\left<N^2\right>_V-\left<N\right>^2_V=4/{\bf b} \left<N\right>_V$
~\cite{Diakonov:1983hh}. Here ${\bf b}=11N_c/3$  is the coefficient of the 1-loop beta function $\beta(\rho_c)\approx {\bf b}/{\rm ln}(\Lambda\rho_c)$ (quenched).  Thus

\be
\left<{\bf Q}^2\right>_V\approx \frac 1{n_I(\tau^2\Delta\eta V_\perp(b)/2)+4/{\bf b}}
\ee
The topological fluctuations are suppressed by the large collision 4-volume. Note that we have ignored the role
of the temperature on the the topological fluctuations in peripheral collisions. Temperature will cause these topological
fluctuations to deplete and vanish at the chiral transition point following the instanton-anti-instanton pairing~\cite{Janik:1999ps}.  So our results will be considered as upper-bounds.

\begin{figure}
\includegraphics[width=65mm]{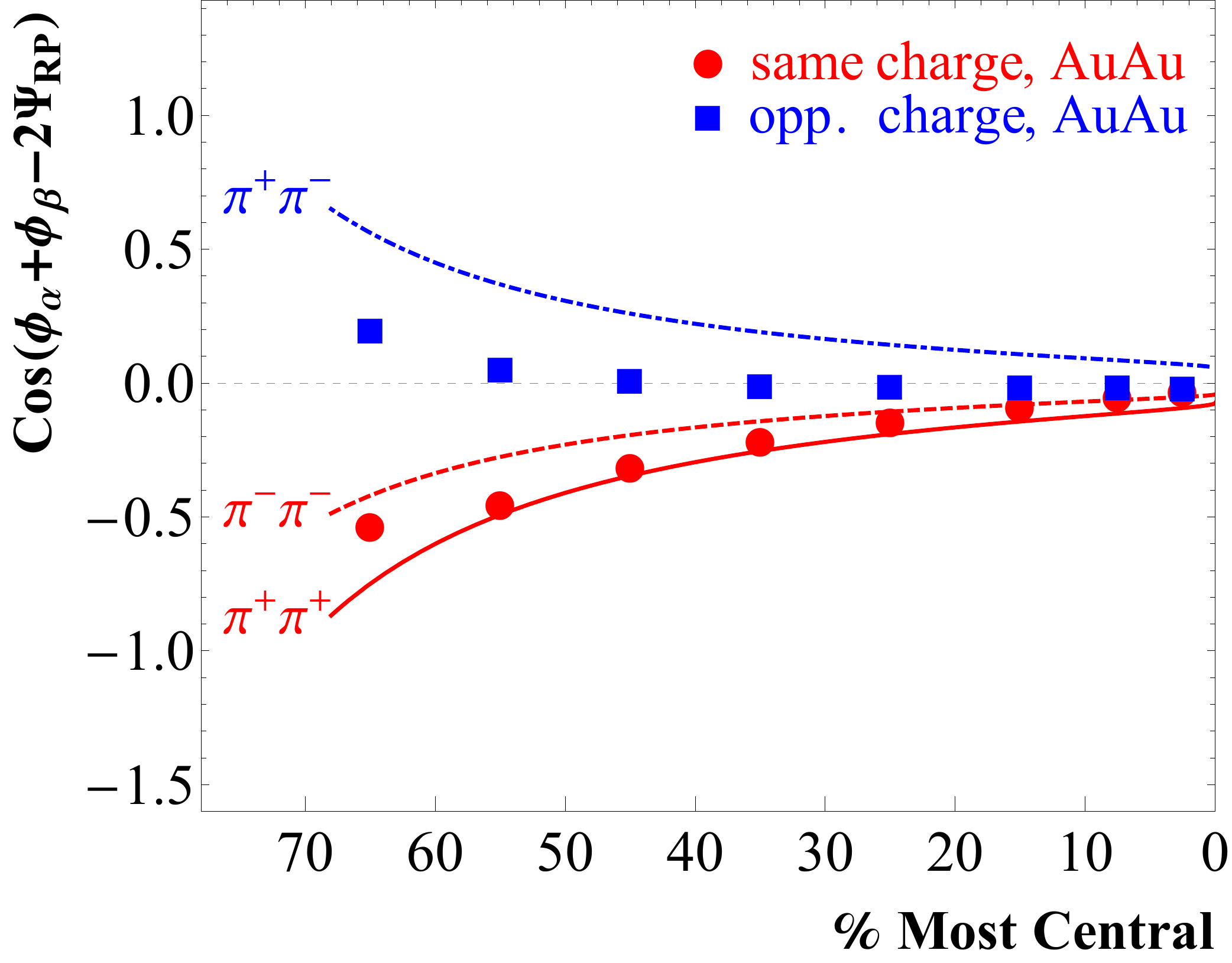}
\caption{Pion azimuthal charge correlations versus the data~\cite{Abelev:2009ac} from STAR  at $\sqrt{s}= 200 {\rm GeV}$. See text.}
\label{auau}
\end{figure}

\begin{figure}
\includegraphics[width=65mm]{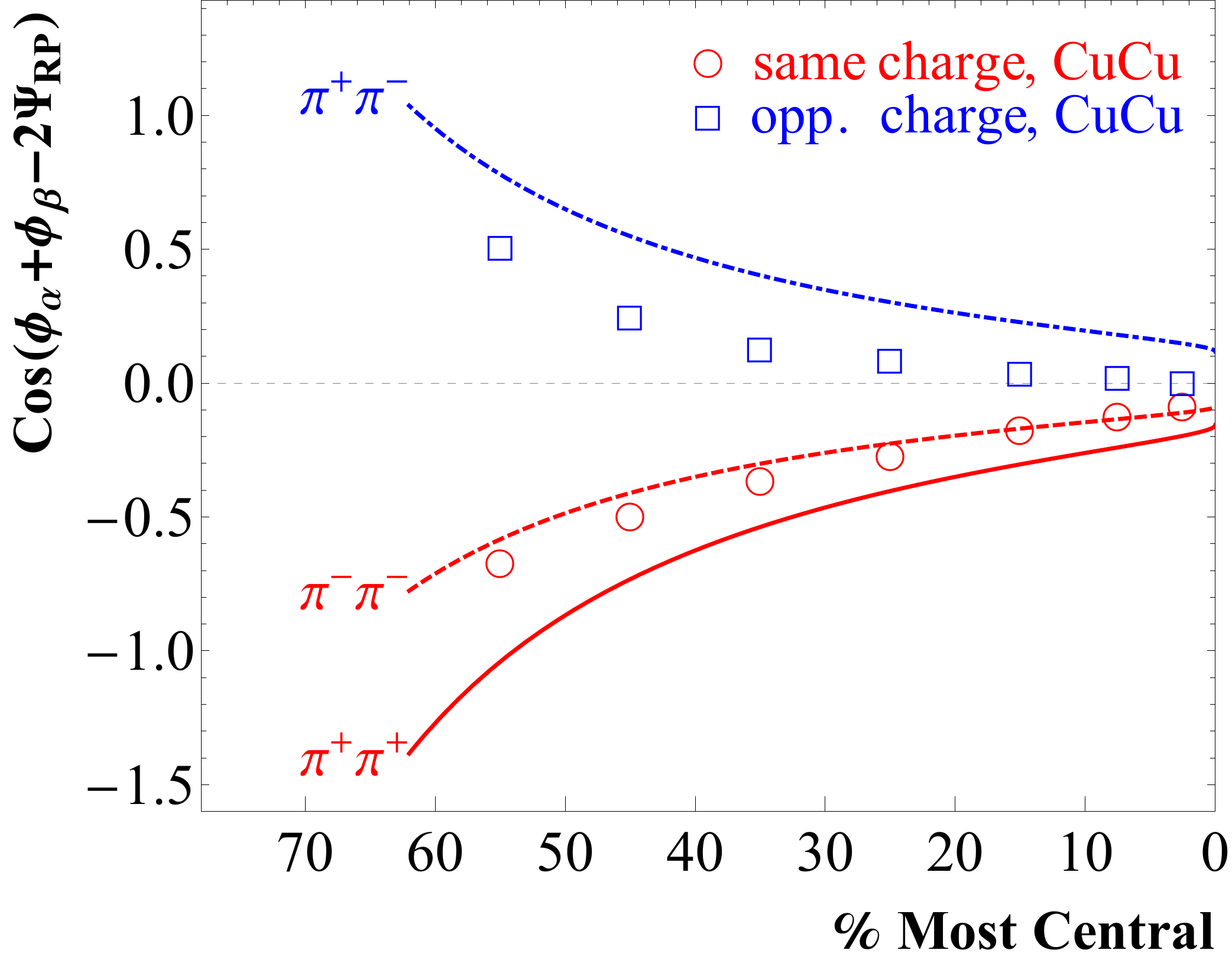}
\caption{Pion azimuthal charge corrlations versus the data~\cite{Abelev:2009ac} from STAR  at $\sqrt{s}= 200 {\rm GeV}$. See text.}
\label{cucu}
\end{figure}

\begin{figure}
\includegraphics[width=65mm]{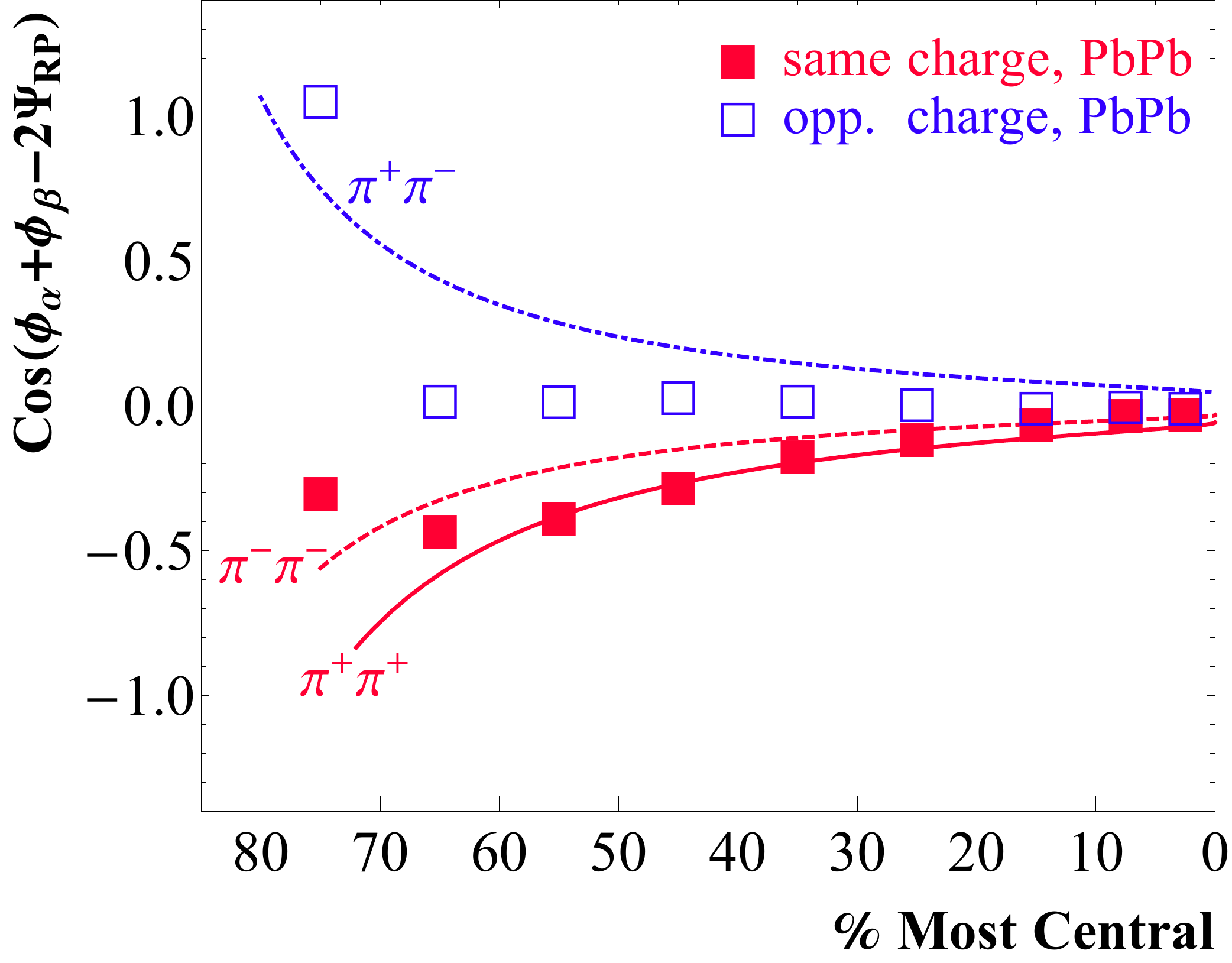}
\caption{Pion azimuthal charge correlations versus the data from ALICE~\cite{Selyuzhenkov:2012py}  at $\sqrt{s}=2.76 {\rm TeV}$. See text.}
\label{pbpb}
\end{figure}

For simplicity, we assume $\left< x\right> = 1/3$ for each parton and $\left< x_F \right> = \left<z \right> = 0.5$.  We also 
fix $\tau = 3 \,{\rm fm}$ to be the maximum duration of the magnetic field polarization, and set
the pseudo-rapidity interval to $( -4, 4)$ for STAR  and $(-5, 5)$ for ALICE. The radius of the colliding nuclei 
will be set to $R=1 {\rm fm} \times \sqrt[3]{A} $ where $A$ is the atomic number. The centrality is approximated as 
$n\% = b^2/(2 R)^2$~\cite{Aguiar:2004uq}. Our results are displayed in Fig.~\ref{auau} for  $AuAu$ and  Fig.~\ref{cucu} for $CuCu$ 
collisions  at $\sqrt{s}=200 {\rm GeV}$ (STAR), and  in Fig.~\ref{pbpb} for $PbPb$ collisions  at $\sqrt{s} = 2.76 {\rm TeV}$
(ALICE).  We recall that~\cite{Voloshin:2004vk}

\be
\left< \cos (\phi_\alpha + \phi_\beta - 2 \Psi_{\rm RP}) \right> \equiv - \left< a_{\alpha} a_{\beta} \right> 
\ee
For the like-charges the results compare favorably with the  data. For the unlike charges they overshoot the  data 
especially for the heavier ion.     Since the magnetic field changes with the impact parameter $b$~\cite{Skokov:2009qp},  it follows that
full proton polarization is only taking place at  $30\%$ and higher centralities. We have checked that the magnetically weighted results with 
the impact parameter do not differ quantitatively from the unweighted results presented in~Figs.~\ref{auau}-\ref{pbpb}.

\section{\label{sec:conclusions}conclusions}

Large chirality flips from  instanton and anti-instanton contributions 
as assessed in polarized $pp$ experiments may contribute substantially to $\mathcal{P}$-odd azimuthal correlations in {\it unpolarized}  $AA$ collisions at RHIC and LHC, thanks to the topological fuctuations in the QCD vacuum and a large induced magnetic field in the prompt part of the collision. The effect is stronger in peripheral collisions and subsede in
central collisions. Simple estimates based on the collision geometry and the magnetic field profile, compare 
 favorably to the currently reported pion azimuthal charge correlations by the STAR and ALICE collaborations.  Our arguments 
involve only polarized protons in the presence of topological fluctuations in the confined vacuum, and therefore complement the chiral magnetic effect suggested in the deconfined vacuum~\cite{Kharzeev:1998kz,Kharzeev:2004ey,Kharzeev:2007jp,Fukushima:2008xe}. 

\section{\label{sec:acknowledgements}acknowledgements}
{\bf Acknowledgements.\,\,}
This work was supported in parts by the US-DOE grant DE-FG-88ER40388.

\bibliography{instantonref}

\end{document}